%% file: Paper.tex
\documentclass[10pt,conference]{IEEEtran}
\input{Preamble}

\begin{document}

\title{Latency-Constrained Encoded Quantum Teleportation with Punctured Codes}

\author{
  \IEEEauthorblockN{
    Mahmoud Saad Abouamer\textsuperscript{\text{$\dagger$}}, Jakob Kaltoft Søndergaard\textsuperscript{\text{$\dagger$}},
    Petar Popovski}
  \IEEEauthorblockA{
    Department of Electronic Systems, Aalborg University,
    9220 Aalborg, Denmark\\
    Email: \{mahmoudabo, jakobks, petarp\}@es.aau.dk}
  \thanks{ This work was supported, in part, by the Danish National Research Foundation (DNRF),
      through the Center CLASSIQUE, grant nr.\ 187. \par \textsuperscript{\text{$\dagger$}} Equal contribution.}
}

\maketitle

\begin{abstract}
Quantum teleportation is a key protocol for transmitting quantum information using entanglement and classical communication. Its reliability is constrained by both the availability and fidelity of shared entangled pairs, which are affected by stochastic generation and memory decoherence. In this work, we focus on \emph{encoded teleportation}, in which quantum information is encoded using a quantum error-correcting code and transmitted as a codeword. We evaluate reliability in terms of logical error probability, considering latency-constrained settings where entanglement is accumulated over time and degrades while in memory. We develop a unified framework that captures the interaction between entanglement availability, decoherence, and coding decisions. Our results show that the benefits of longer codes depend  on the availability and fidelity of entangled pairs, as acquiring additional resources introduces delays that can reduce their quality. To address this latency-reliability tradeoff, we leverage code puncturing to enable flexible encoded teleportation, allowing the effective code length to adapt across different latency regimes while preserving a common stabilizer structure. Numerical results show that encoded teleportation can provide substantial reliability gains over uncoded transmission under a common entanglement-acquisition latency constraint, and that selecting appropriate punctured codes improves performance across varying latency budgets. Overall, our results highlight the importance of resource-aware adaptation for reliable quantum networking.
\end{abstract}

\begin{IEEEkeywords}
Quantum networks, entanglement distribution, quantum error correction, encoded teleportation, code puncturing.
\end{IEEEkeywords}

\section{Introduction}
Quantum networks enable the distribution of quantum information across different quantum processors, supporting applications such as distributed quantum computation~\cite{Quant_net_d_computing_1,Quant_net_d_computing_2}. \emph{Quantum teleportation} serves as a fundamental primitive for such networks, enabling the transfer of quantum states through shared entanglement and classical communication~\cite{Quant_net_d_computing_2}. The reliability of teleportation is therefore fundamentally tied to the quality of the shared entanglement.

To improve teleportation reliability, one direct approach is to improve the fidelity of the shared entangled pairs before they are used for teleportation. This is commonly achieved through entanglement purification, where multiple imperfect entangled pairs are processed to obtain fewer pairs with improved fidelity using local operations and classical communication~\cite{Deutsch1996DEJMPS}. However, purification introduces additional processing, coordination, and storage overhead, during which stored entanglement decoheres~\cite{pur_with_latency_arxiv,pur_with_latency_pre_conf}. Consequently, practical latency and storage constraints can limit the fidelity that can be achieved through resource-level optimization alone.

This motivates \emph{encoded teleportation}, where reliability is improved beyond individual EPR pair fidelity by encoding quantum information using a quantum error-correcting code prior to teleportation~\cite{Lorenzo2024,resource-adaptive_teleportation}. A logical qubit is encoded across multiple physical qubits, each teleported using an independent entangled pair, followed by decoding to detect and correct errors introduced by imperfect entanglement. While logical-level protection can improve reliability beyond individual EPR pair fidelity, it also increases the number of entangled resources required for communication. In particular, an $[[n,k]]$ code requires $n$ entangled pairs to transmit $k$ logical qubits. While longer codes improve error-correction capability, they also increase resource consumption and latency. However, these works treat entanglement quality as independent of code length and do not account for how acquiring larger entanglement resources can alter their fidelity through waiting-time variability and memory decoherence.

As illustrated in Fig.~\ref{fig:latency_memory_tradeoff}, the impact of longer codes in quantum networks extends beyond the local encoding and decoding operations and includes the time required to acquire entanglement resources. Since encoded teleportation requires multiple entangled pairs to transmit a logical state, the network must first accumulate these resources before the encoded teleportation can be completed. Such collections of entangled pairs, referred to as \emph{entanglement packets}~\cite{davies2024tools}, are generated probabilistically, making the acquisition time variable. {Under latency constraints, the network may need to operate at higher entanglement generation probabilities to meet the required timing, which can result in lower-fidelity entangled pairs~\cite{davies2024tools}.} In addition, pairs generated earlier must be stored while the remaining pairs are acquired, during which they decohere and degrade in fidelity.

\input{Figures/figure_0}

Consequently, increasing the packet size to support a longer code reduces the quality of the entanglement used for teleportation. In some regimes, this reduction in pair quality can outweigh the  error-correction gain of a longer code, so longer codes do not necessarily improve reliability. This effect is driven by probabilistic acquisition delays and subsequent memory decoherence, which has no direct analogue in conventional classical communication resources. The resulting latency-reliability tradeoff, illustrated in the contrasting regimes of Fig. \ref{fig:latency_memory_tradeoff}, depends on both the number and quality of available entangled pairs. This work focuses on network-related imperfections arising from stochastic entanglement generation, storage-induced decoherence, and noisy shared entangled pairs.

The latency-reliability tradeoff affects the optimized choice of code length depending on the network conditions. As illustrated conceptually in Fig.~\ref{fig:decision_regions}, evaluating the logical error probability under different latency constraints can reveal distinct decision regions separated by crossover boundaries. Although a given code length can, in principle, be operated under different latency targets by adjusting the entanglement generation probability, doing so changes the quality of the generated entangled pairs and therefore the resulting reliability. Thus, different effective code lengths can minimize the logical error probability in different operating regimes, leading to thresholds where the preferred code size changes. Since entanglement generation, link quality, and latency requirements can vary across operating conditions, a fixed code choice may be suboptimal. We investigate these tradeoffs to identify decision regions and motivate a latency-aware code-selection strategy.

To support this, we adopt \emph{puncturing of quantum error-correcting codes} \cite{gundersen2025puncturing} as a mechanism for modifying code length and distance by selectively removing physical qubits.  This makes puncturing a practical mechanism for adaptive encoded teleportation driven by resource availability. In particular, nodes can operate within a common base-code architecture and dynamically use only the number of entangled pairs required by the chosen punctured code, enabling adaptation to resource availability while reducing the need to switch between unrelated encoding and decoding implementations. Since shared entangled pairs are consumed as network communication resources, whereas encoding and decoding are performed locally, puncturing can reduce the number of entangled pairs required for a teleportation request while maintaining a common base-code structure. Moreover, coding gains are obtained only at certain punctured code lengths rather than continuously with packet size.  We refer to these discrete operating points as \emph{puncturing tiers}, which create non-trivial trade-offs between the latency required to acquire larger entanglement packets and the reliability gains provided by stronger error correction.
  
While~\cite{resource-adaptive_teleportation} considered punctured encoded teleportation under fixed entanglement availability and quality, independent of code length, this work incorporates the entanglement-acquisition process into the code-selection problem. Since an $[[n,k]]$ code requires $n$ entangled pairs, increasing the code length can improve error protection but can also increase acquisition time, memory decoherence, and fidelity heterogeneity across the packet. As a result, the best puncturing tier depends on the latency and link-quality regime, providing a network-aware rationale for when different punctured codes should be used.

More broadly, this work bridges entanglement generation and its use in scalable quantum network architectures. Here, entanglement generation refers to the successful establishment of shared entanglement between communicating nodes. In the entanglement generation literature (e.g.,~\cite{davies2024tools}), the focus is on generating and maintaining entanglement, often without considering how it is ultimately used in applications such as teleportation. Conversely, encoded-teleportation studies often abstract away the stochastic acquisition and storage of entanglement resources. We investigate this interface by evaluating how stochastic entanglement availability affects logical teleportation reliability and how punctured codes can adapt to resource constraints. Such adaptiveness is particularly relevant for early-generation quantum access networks, where users exhibit diverse quality-of-service requirements and hardware capabilities vary significantly. In these settings, a common base encoding, shared across a network domain such as a wireless quantum cell~\cite{popovski20251q}, can be adaptively punctured to accommodate individual users. This provides a potential architectural approach for balancing performance, complexity, and resource efficiency.

The main contributions of this work are as follows:
\begin{itemize}
    \item We develop a unified framework for latency-constrained encoded teleportation that jointly models stochastic entanglement generation, memory decoherence, and punctured quantum error-correcting codes.
    \item We investigate the latency-reliability tradeoff by comparing uncoded teleportation and multiple encoded teleportation schemes under a common average entanglement-acquisition latency constraint, highlighting how latency, entanglement fidelity, and logical error probability are coupled under resource constraints.
    \item Through numerical simulations, we investigate latency- and link-dependent decision regions in which different puncturing tiers minimize the logical error probability, demonstrating the potential of adaptive puncturing-based code selection to outperform fixed code-selection strategies.
\end{itemize}

\input{Figures/Figure_4}

\section{System Model}
In this section, we introduce a unified model for latency-constrained encoded teleportation based on stochastic entanglement generation. The model captures the interaction between entanglement generation, storage-induced decoherence, and quantum error correction, enabling a cross-layer analysis of latency-reliability trade-offs in quantum networks. Throughout this work, entanglement generation refers to the successful generation and distribution of an EPR pair shared between the communicating nodes.

As illustrated in Fig.~\ref{fig:latency_memory_tradeoff}, encoded teleportation proceeds through a sequence of stages including entanglement generation, encoding, teleportation, and decoding. Since encoding and decoding are local operations and classical communication during teleportation can be performed in parallel, we focus on the latency associated with entanglement generation, which is typically the dominant contribution in many physical implementations. In particular, the time required to generate an entanglement packet of $n$ pairs determines the overall latency. Since the packet size $n$ governs both the acquisition time and the strength of error correction, this induces a trade-off between resource availability, fidelity, and logical reliability. In particular, increasing $n$ improves error-correction capability but requires longer waiting times to accumulate entanglement, leading to additional decoherence and reduced fidelity.

The goal is to select the code length $n$ that minimizes the logical error probability under a latency constraint. To accommodate varying operating conditions, we consider puncturing of a base code, allowing the effective code length to be adapted while preserving a common stabilizer structure. This provides a structured way to realize different latency-reliability tradeoffs without switching among unrelated code families, as detailed in the following subsections.

\subsection{Encoded Teleportation}
\subsubsection{Imperfect Teleportation} Teleportation using ideal Bell pairs and perfect local operations faithfully transmits an arbitrary quantum state. In realistic settings, however, the shared entangled resource is imperfect, and the protocol can be interpreted as a Pauli channel whose error probabilities $\{p_I, p_Z, p_X, p_Y\}$ are determined by the Bell-state decomposition of the resource~\cite{Lorenzo2024}:
\begin{align}
\rho &= p_I \ket{\Phi^+}\bra{\Phi^+} 
     + p_Z \ket{\Phi^-}\bra{\Phi^-} \nonumber\\
     &\quad + p_X\ket{\Psi^+}\bra{\Psi^+} 
     + p_Y\ket{\Psi^-}\bra{\Psi^-},
     \label{eq:teleportation_pauli_channel}
\end{align}
where $p_I+p_Z+p_X+p_Y=1$.

\subsubsection{Logical Reliability under CSS Codes}
To mitigate these errors, we employ encoded teleportation using CSS codes. Instead of transmitting a single qubit, the state is encoded into an $[[n,k,d_Z,d_X]]$ CSS code, requiring an entanglement packet of $n$ EPR pairs, one for each physical qubit. Due to the stochastic generation process (see Sec.~\ref{section:entanglement_generation}) and storage-induced decoherence (see Sec.~\ref{section:memory decoherence}), these pairs generally have non-identical fidelities, resulting in heterogeneous Pauli error probabilities $({p_{I,i},}p_{Z,i}, p_{X,i}, p_{Y,i})$ across qubits.

CSS codes are well-suited to Pauli noise as they decouple $X$ and $Z$ error correction. Each branch can therefore be treated as a binary symmetric channel with effective error probabilities~\cite{CSS_Error_Prob}
\begin{equation}
q_{X,i} = p_{X,i} + p_{Y,i}, \qquad
q_{Z,i} = p_{Z,i} + p_{Y,i},
\end{equation}
corresponding to bit-flip and phase-flip errors, respectively. This model is adopted from~\cite{CSS_Error_Prob} to evaluate logical reliability across different puncturing tiers.

Furthermore, to capture realistic noise processes, we consider both symmetric and asymmetric Pauli channels. The latter is defined by the ratio
\begin{equation}
\eta = \frac{p_{Z,i}}{p_{X,i}},
\end{equation}
allowing phase-flip errors to occur more frequently than bit-flip errors~\cite{ioffe2007asymmetric}.

Let $t_Z=\lfloor(d_Z-1)/2\rfloor$ and $t_X=\lfloor(d_X-1)/2\rfloor$ denote the correction radii of the code. Due to heterogeneous error probabilities across qubits, errors are independent but not identically distributed, and the number of errors in each branch therefore follows a Poisson-binomial distribution. The probabilities of successful decoding of $X$ and $Z$ errors are given by
\begin{align}
    P_{\mathrm{succ},X} &= \sum_{j=0}^{t_X}
    \sum_{\substack{S \subseteq \{1,\dots,n\}\\ |S|=j}}
    \prod_{i\in S} q_{X,i}
    \prod_{i\notin S} \big(1-q_{X,i}\big), \\
    P_{\mathrm{succ},Z}
    &= \sum_{j=0}^{t_Z}
    \sum_{\substack{S \subseteq \{1,\dots,n\}\\ |S|=j}}
    \prod_{i\in S} q_{Z,i}
    \prod_{i\notin S} \big(1-q_{Z,i}\big).
\end{align}
The resulting logical error probability is
\begin{equation}
    P_L = 1 - P_{\mathrm{succ},X} P_{\mathrm{succ},Z}.
    \label{eq:PL_poisson_binomial}
\end{equation}
The logical reliability above depends on the error probabilities induced by the shared entangled pairs. In practice, these probabilities are determined by how entanglement is generated and stored over time. As entanglement generation is stochastic and dominates the overall latency, the time required to accumulate an entanglement packet directly impacts both the availability and quality of the pairs used for teleportation. 

This establishes a trade-off between latency and entanglement fidelity. As illustrated in Fig.~\ref{fig:decision_regions}, this trade-off gives rise to distinct operating regimes in which different coding strategies are optimal, depending on the latency constraint. We analyze this behavior by modeling entanglement generation and memory decoherence in the following subsections. Furthermore, as the required packet size $n$ directly affects this trade-off, adaptive code selection via puncturing tiers (see Sec.~\ref{section:puncturing tiers}) becomes essential.

\subsection{Entanglement Generation Rate and Fidelity}
\label{section:entanglement_generation}
Entanglement generation is modelled as a discrete-time stochastic process, where in each time slot an attempt is made to generate and distribute an EPR pair between the communicating nodes. {The duration of a slot depends on the implementation, with experimental demonstrations typically reporting entanglement-generation rates on the order of 10–20 kHz}\cite{zhang2024fast, craddock2024automated}.

The entanglement generation and storage process is illustrated in Fig.~\ref{fig:age_attempts}, where successful generation attempts occur over time and are stored until a complete entanglement packet has been successfully distributed.

We model entanglement generation as a sequence of independent Bernoulli trials, where each attempt succeeds with probability $p$. When the success probability of a single attempt is very small, the sender may perform a batch of $M$ attempts within a single time slot, in which case $p$ corresponds to the probability that at least one of the attempts succeeds.

The success probability is inherently related to the fidelity of the generated pair; protocols that aim for higher fidelity typically achieve lower generation rate. Following \cite{davies2024tools}, we model this trade-off by relating the initial fidelity $F_0$ of a successfully generated pair to the generation probability as
\begin{equation}
\label{eq:F0}
    F_0(p)=1-\frac{1-(1-p)^{1/M}}{3p_d},
\end{equation}
where $M$ denotes the batch size and $p_d$ is the photon detection probability{, which captures physical link characteristics such as photon collection efficiency and detector geometry that can vary with the spatial configuration of the link and across users~\cite{photon_Det_exp}.}

The fidelity $F_0(p)$ characterizes the fidelity of entangled pairs at the time of generation. However, these pairs are stored in quantum memory while additional entanglement is accumulated. During this time, qubits are subject to decoherence, causing their fidelities to degrade as illustrated in Fig.~\ref{fig:age_attempts}. In the example shown in Fig.~\ref{fig:age_attempts}, a packet of three entangled pairs is requested, but completing the packet requires five generation attempts. Since successful pairs must be stored while waiting for the remaining pairs, earlier successes accumulate age and gradually decohere. Consequently, when the final pair is generated (Age 0), previously generated pairs may have aged (e.g., Ages 2 and 4), leading to heterogeneous fidelities within the completed packet.

\begin{figure}[t]
\centering    
\resizebox{1.0\linewidth}{!}{\input{Figures/age_Time}}
\caption{Example of the generation process with entanglement packet size $n=3$. Generated EPR pairs are stored until a complete packet is formed. Their fidelity degrades over time due to decoherence, resulting in heterogeneous fidelities across the packet.}
\label{fig:age_attempts}
\end{figure}

\subsection{Memory Decoherence}
\label{section:memory decoherence}
We model the degradation as exponential decay such that a qubit stored for $t$ time slots has fidelity
\begin{equation}
\label{eq:decoherence}
    F_t=\frac{1}{2}+\left(F_0-\frac{1}{2}\right)e^{-t/T},
\end{equation}
where $T$ is the coherence time of the quantum memory.

To mitigate excessive degradation caused by long storage times, we impose a finite storage window of size $w$, such that entangled pairs are discarded if not used within $w$ time slots similarly to \cite{davies2024tools}. On one hand, this prevents severely decohered pairs from being used in teleportation to reduce the impact of excessively aged pairs on the logical error probability. On the other hand, it constrains the time available to accumulate an entanglement packet. Consequently, the finite window size introduces an inherent trade-off between resource availability and fidelity. The trade-off is further enhanced by latency requirements, which determine how long the communicating nodes can wait to form an entanglement packet.

Together, the entanglement generation model, memory decoherence, and encoded teleportation framework define the error characteristics and reliability of the transmitted quantum information under realistic network conditions.

\section{Latency-Constrained Encoded Teleportation and Adaptive Coding}
In this section, we investigate how latency constraints shape entanglement availability and fidelity, and how these in turn impact the performance of encoded teleportation. In particular, the time required to generate an entanglement packet couples resource availability with decoherence, creating a trade-off between latency and logical reliability. This trade-off directly influences the choice of code length $n$, motivating adaptive coding strategies such as puncturing, which are presented in this section.

\subsection{Latency Requirement and Waiting Time}
\label{section:Latency Requirement and Waiting Time}
We impose a latency requirement $L$, specifying that an entanglement packet must be successfully generated within an average of $L$ time slots, reflecting application-level quality of service constraints. For a given packet size $n$ and storage window $w$, this latency constraint determines the required entanglement generation probability $p$.

Let $\mathbb{E}[L(w,n)]$ denote the expected waiting time until an entanglement packet of size $n$ is successfully generated within a window of size $w$. In the special case where $n=1$, {the waiting time} follows a geometric distribution with success probability $p$, thus $\mathbb{E}[L(w,1)]=1/p$. However, for larger $n$, obtaining the exact expected waiting time requires solving a linear system of dimension $\binom{w-1}{n-1}$, which quickly becomes computationally intractable \cite{davies2024tools}. Instead, for $n>1$, we approximate the expected waiting time as
\begin{equation}
\label{eq:waiting_time}
    \mathbb{E}[L(w,n)]\approx\left(\frac{1}{P_e}-1\right)\left(w+\frac{n}{2p}\right)+\frac{n}{p},
\end{equation}
where
\begin{equation}
    P_e=\sum_{\ell=n-1}^w \binom{w}{\ell}p^\ell(1-p)^{w-\ell}
\end{equation}
is the probability that at least $n-1$ additional successes occur within a window of $w$ attempts following the first success.

Using this approximation, we determine the generation probability $p$ required to satisfy the latency constraint by solving $\mathbb{E}[L(w,n)]=L$ numerically. Comparison with exact evaluations shows that \eqref{eq:waiting_time} conservatively overestimates the expected waiting time, and the approximation becomes increasingly accurate as $p \to 1$. Consequently, satisfying a given latency constraint requires a slightly higher generation probability $p$ than under the exact model, introducing a conservative bias: latency requirements are more likely to be met, but at the cost of reduced initial entanglement fidelity due to the coupling between generation rate and fidelity in \eqref{eq:F0}.

\begin{figure}
    \centering
    \includegraphics[width=\linewidth]{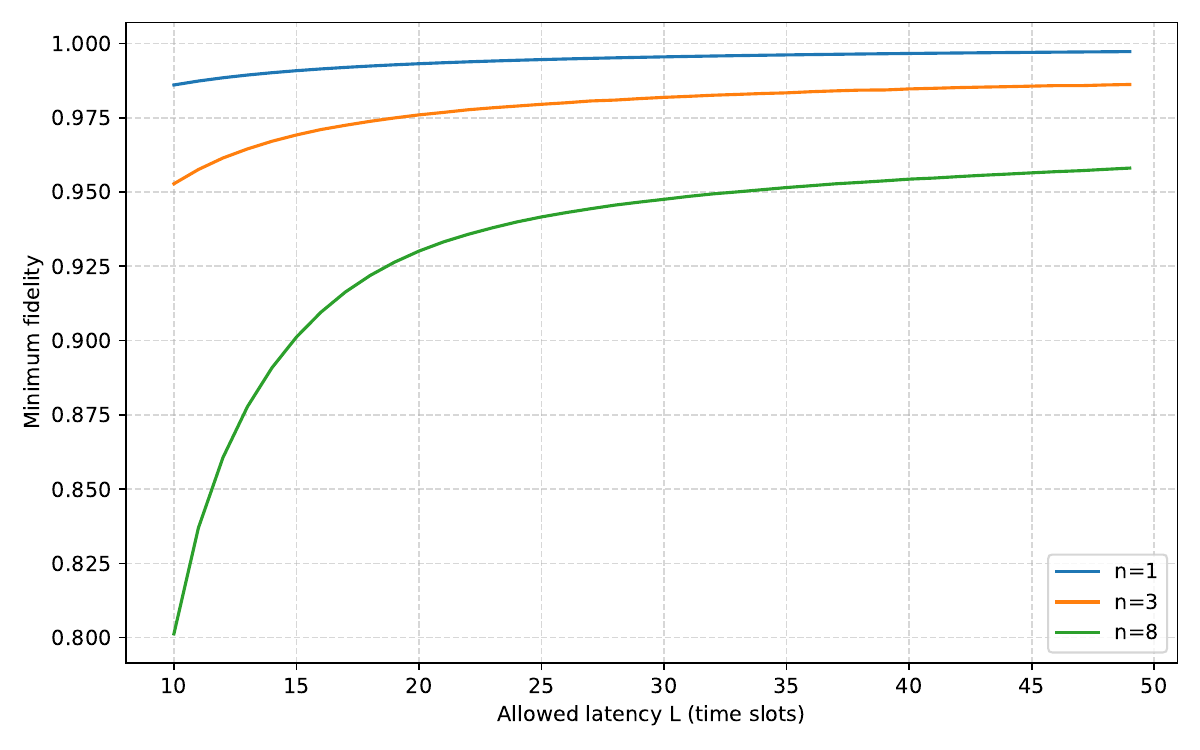}
    \caption{Minimum fidelity of the entangled pairs as a function of the latency constraint $L$. The minimum corresponds to the oldest stored pair at the time the packet is completed. Allowing larger latency enables higher-quality entanglement as pairs can be generated with lower success probability and thus higher initial fidelity $F_0$.}
    \label{fig: min_fidelity_vs_latency}
\end{figure}

Since the generation probability is physically coupled to the quality of the generated entanglement, this in turn determines the initial fidelity through $F_0=F_0(p)$, establishing a direct trade-off between latency constraint and entanglement quality. To illustrate this latency-fidelity coupling, we evaluate the minimum fidelity of the entangled pairs, corresponding to the oldest stored pair in a packet, used for encoded teleportation as a function of the latency constraint $L$, as shown in Fig.~\ref{fig: min_fidelity_vs_latency}.

This trade-off plays a central role in determining which coding strategy is preferred under different latency regimes. As shown in Fig.~\ref{fig: min_fidelity_vs_latency}, longer codes require more aggressive entanglement generation rates to satisfy a given latency constraint, which reduces the fidelity of the generated pairs. Consequently, the system must balance using a small number of high-fidelity qubits, corresponding to uncoded or short-code teleportation, against using a larger number of lower-fidelity qubits that provide redundancy through error correction. Determining where the reliability gain from coding outweighs the loss in entanglement quality depends jointly on latency constraints and link conditions. This motivates adaptive coding strategies that select the appropriate coding policy for the current operating regime.

\subsection{Puncturing Tiers}
\label{section:puncturing tiers}

To adapt to latency-constrained entanglement generation, we consider puncturing of quantum error-correcting codes, which allow the code length to be reduced by selectively removing qubits while retaining a common encoding structure at the cost of lowering code distance \cite{gundersen2025puncturing}. In the context of encoded teleportation, this enables different entanglement packet sizes to be considered under latency constraints, at the cost of reduced error-correction capability.

Puncturing a base code can produce many candidate code lengths. However, multiple punctured variants may provide the same error-correction capability, characterized by identical $(d_Z,d_X)$ values. Since larger $n$ requires more entangled pairs and incurs higher latency, we retain only the smallest-$n$ code for each distinct $(d_Z,d_X)$ pair. This results in a discrete set of operating points, referred to as \textit{puncturing tiers}. In this work, we consider four tiers derived from the same length-$17$ CSS base code following \cite{resource-adaptive_teleportation}:
\begin{equation*}
    [[17,1,5,5]],\quad [[13,1,5,3]],\quad [[8,1,3,3]],\quad \text{uncoded}.
\end{equation*}
These tiers preserve a single logical qubit ($k=1$) while offering different packet sizes and error-correction capabilities. In particular, the $[[13,1,5,3]]$ tier provides stronger protection against phase errors than bit-flip errors, making it well suited for asymmetric noise regimes where phase errors dominate \cite{ioffe2007asymmetric}. The final tier corresponds to uncoded teleportation using a single entangled pair.

Each tier therefore represents a different trade-off between entanglement packet size, acquisition latency, and error-correction capability. While larger codes generally provide stronger error correction for equal fidelities, they require larger entanglement packets and longer acquisition times, which reduce the fidelity of stored pairs due to decoherence.

This interplay between stochastic entanglement generation, storage-induced decoherence, and coding choices defines the trade-off studied in this work and motivates the need for adaptive encoded teleportation strategies.

\subsection{Offline Optimization and Online Deployment}
In this section, we present an operational use of the proposed framework, illustrated in Fig.~\ref{fig:system_model}.  The framework consists of two phases for adaptive encoded teleportation. In the first phase, system parameters, including generation statistics, link quality, and memory coherence times, are evaluated offline across candidate puncturing tiers to analyze the trade-offs between logical error probability, latency, and link quality. For each relevant operating point, such as latency requirement or target reliability, the logical error probability is computed for all candidate codes, from which an optimized coding policy can be extracted. {In particular}, for a given operating point defined by the latency constraint $L$, detection probability $p_d$, and noise asymmetry $\eta$, the selected puncturing tier is
\begin{equation}
n^{\star}(L,p_d,\eta)=
\arg\min_{n \in \mathcal{N}}
P_L(n;L,p_d,\eta),
\end{equation}
where $\mathcal{N}=\{1,8,13,17\}$ denotes the set of candidate puncturing tiers. The resulting policy can be stored as a precomputed policy map.

In the second phase, the framework is deployed during network operation. When a teleportation request arrives, the associated quality-of-service requirements and operating conditions are used to determine the appropriate puncturing tier from the precomputed policy map. The required entanglement packet is then requested. {After successful generation, the data qubit is encoded using the selected punctured code, and the resulting physical qubits are teleported using the entanglement packet and decoded at the receiver.}

The numerical results in Sec.~\ref{section:numerical_results} focus on the first phase to investigate these trade-offs and motivate the resulting policy map.
\input{Figures/figure_5}

\section{Numerical Results}
\label{section:numerical_results}

In this section, we evaluate the performance of latency-constrained encoded teleportation, with the goal of quantifying the trade-off between latency and logical reliability, and comparing different coding strategies under a common latency constraint. To enable this comparison, we follow the modeling framework developed in Sec.~\ref{section:entanglement_generation}--Sec.~\ref{section:memory decoherence} and Sec.~\ref{section:Latency Requirement and Waiting Time}. For a given latency constraint $L$ and code length $n$, we determine the entanglement generation probability $p$ by solving $\mathbb{E}[L(w,n)] = L$ using the approximation in \eqref{eq:waiting_time}. This ensures that all schemes satisfy the same latency constraint on average. The resulting value of $p$ determines the initial fidelity of generated entangled pairs through $F_0(p)$ in \eqref{eq:F0}.

We then simulate the stochastic entanglement generation and storage process described in Sec.~\ref{section:entanglement_generation} and Sec.~\ref{section:memory decoherence} until a successful entanglement packet of the corresponding size is obtained. Because successful pairs are generated at different time instants, the qubits in a packet experience different storage durations prior to teleportation. This results in heterogeneous fidelities across the $n$ qubits, which are computed using \eqref{eq:decoherence}. The resulting fidelities are then mapped to qubit-wise Pauli error probabilities $({p_{I,i},}p_{Z,i}, p_{X,i}, p_{Y,i})$ according to the teleportation channel model in \eqref{eq:teleportation_pauli_channel}. Specifically, for a qubit with fidelity $F_i$, we set $p_{I,i}=F_i$ and distribute the remaining probability mass $1-F_i$ across Pauli errors according to the underlying noise model: under symmetric noise $p_{X,i}=p_{Y,i}=p_{Z,i}=(1-F_i)/3$, while under asymmetric noise $p_{X,i}=p_{Y,i}$ and $\eta = p_{Z,i}/p_{X,i}$, with all probabilities satisfying $p_{I,i}+p_{Z,i}+p_{X,i}+p_{Y,i}=1$.

Using these heterogeneous error probabilities, we evaluate encoded teleportation under CSS codes corresponding to the puncturing tiers described in Sec.~\ref{section:puncturing tiers}. For each Monte Carlo realization, we simulate the stochastic entanglement-generation and storage process to obtain the waiting times and storage ages of the qubits in the entanglement packet, which determine their heterogeneous fidelities. These fidelities are then mapped to qubit-wise Pauli error probabilities, and the corresponding logical error probability $P_L$ is computed using \eqref{eq:PL_poisson_binomial}. The reported results are obtained by averaging $P_L$ over $10^5$ independent realizations.

{For our numerical results, we use a coherence time of $T=10000$ time slots. Since the proposed framework is independent of the physical duration of a time slot, the coherence time should be interpreted relative to the underlying implementation. Relative to the representative time-slot durations discussed in Sec.~\ref{section:entanglement_generation}, this corresponds to a coherence time {on} the order of one second, consistent with reported coherence times of  long-lived quantum memories~\cite{simon2010quantum}.}

This procedure enables a comparison across coding schemes, as each operates under the same average latency constraint while capturing the coupling between entanglement generation rate, fidelity, and logical reliability. In particular, longer codes require higher generation probabilities to satisfy the latency constraint, which reduces the fidelity of the generated entanglement and can offset the benefits of increased error-correction capability.

\subsection{Impact of Average Latency Constraint $L$}

\begin{figure}[t]
\centering    
\includegraphics[width=\linewidth]{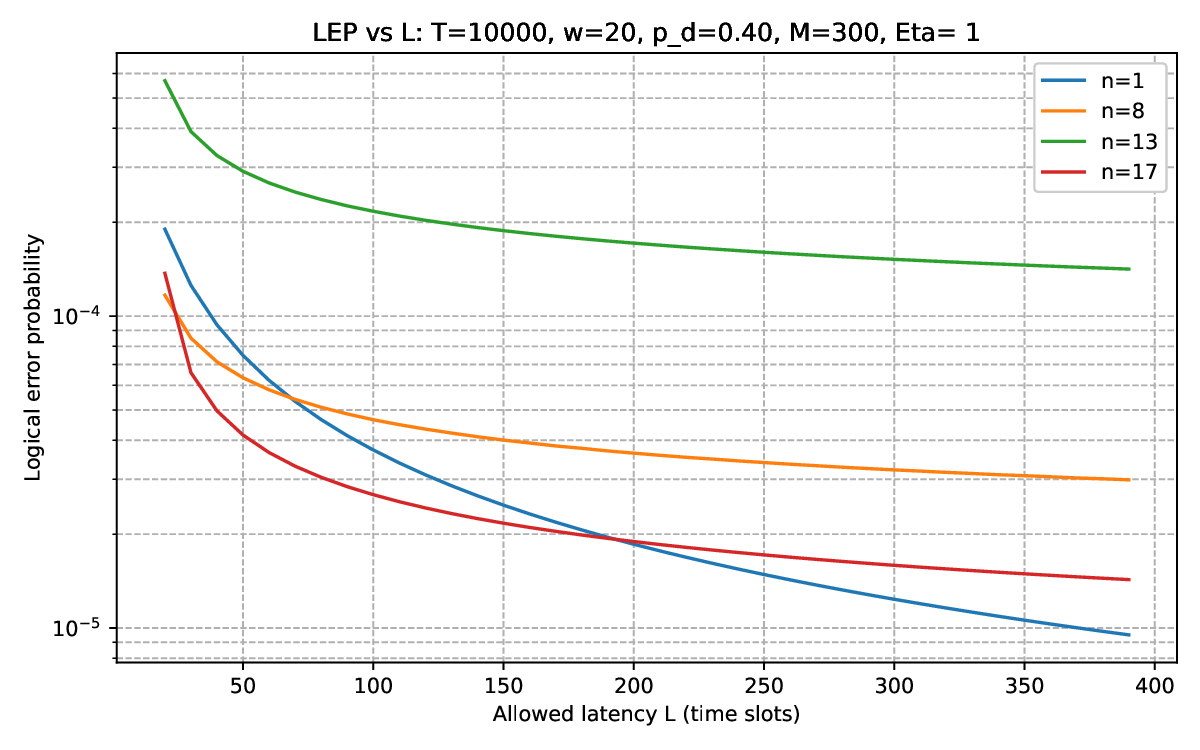}
\caption{Logical error probability $P_L$  as a function of allowed latency $L$ for different puncturing tiers $n$ under symmetric errors.}
\label{fig:pl_vs_l_symm}
\end{figure}
\begin{figure}[t]
\centering    
\includegraphics[width=\linewidth]{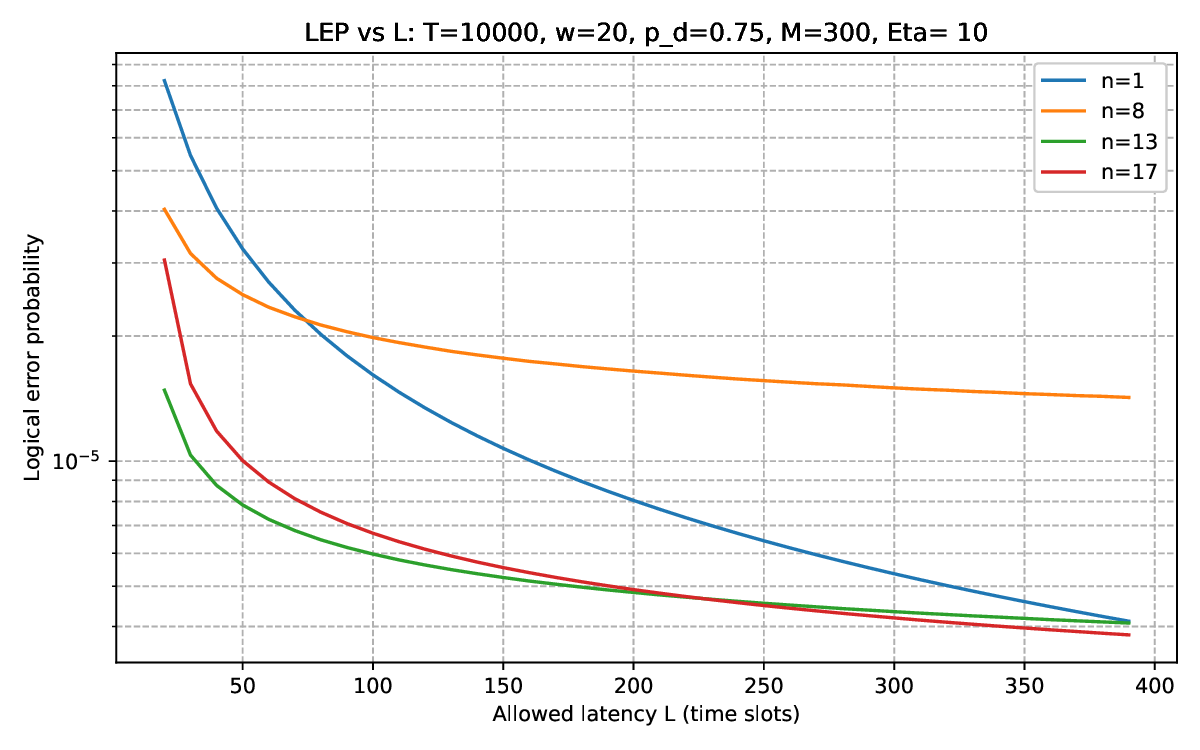}

\caption{Logical error probability $P_L$ as a function of latency constraint $L$ for different puncturing tiers $n$ under asymmetric errors ($\eta=10$).}
\label{fig:pl_vs_l_asymm}
\end{figure}
We evaluate the logical error probability $P_L$ as a function of the latency constraint $L$ for different puncturing tiers, corresponding to different code lengths $n$. This allows us to demonstrate how an optimized coding strategy depends on the available latency budget, and to identify regimes in which encoded teleportation outperforms uncoded transmission, as well as regimes where adaptive coding across puncturing tiers provides significant performance gains over fixed coding strategies.

In Fig.~\ref{fig:pl_vs_l_symm}, we show the logical error probability $P_L$ as a function of the latency constraint $L$ for different coding tiers $n$, with $n=1$ corresponding to the uncoded case. Increasing the allowed latency budget improves entanglement quality, reducing logical error probabilities across all transmission strategies. Under tight latency constraints, encoded teleportation outperforms uncoded transmission, with distinct decision regions in which the $n=8$ punctured code (briefly) and the $n=17$ code respectively achieve the lowest logical error probability. In this regime, the fidelity of a single entangled pair is too low for reliable uncoded transmission, and error correction through encoding reduces logical error probability despite requiring teleportation across multiple physical qubits. However, beyond approximately $L=200$, uncoded transmission becomes favorable, as the fidelity of individual entangled pairs becomes sufficiently high that physical error probabilities are already very small. In this regime, encoded teleportation requires transmitting multiple physical qubits through independent teleportation channels, and the additional exposure to physical errors can outweigh the coding gains of larger codes. These decision regions demonstrate that puncturing enables adaptive selection across coding tiers under varying latency constraints.

In Fig.~\ref{fig:pl_vs_l_asymm}, we show that under asymmetric errors with $\eta=10$, the role of the asymmetrically punctured $n=13$ code becomes pronounced, achieving the lowest logical error probability over most of the latency range. In contrast to the symmetric case, encoded teleportation remains favorable for nearly all values of $L$, with uncoded transmission and the $n=17$ code only catching up briefly at the largest allowed latencies. These results show that asymmetry can significantly shift the best-performing coding tier and further enhance the benefits of adaptive puncturing. Moreover, under resource constraints, encoded teleportation provides a significant reduction in logical error probability (corresponding to improved logical reliability) across nearly the entire latency regime considered. 

Taken together, the symmetric and asymmetric results demonstrate that adaptive puncturing can accommodate distinct error environments while maintaining a common stabilizer structure and flexibly selecting coding tiers matched to the operating regime.

\subsection{Impact of Detection Probability $p_d$}

\begin{figure}[t]
\centering    
\includegraphics[width=\linewidth]{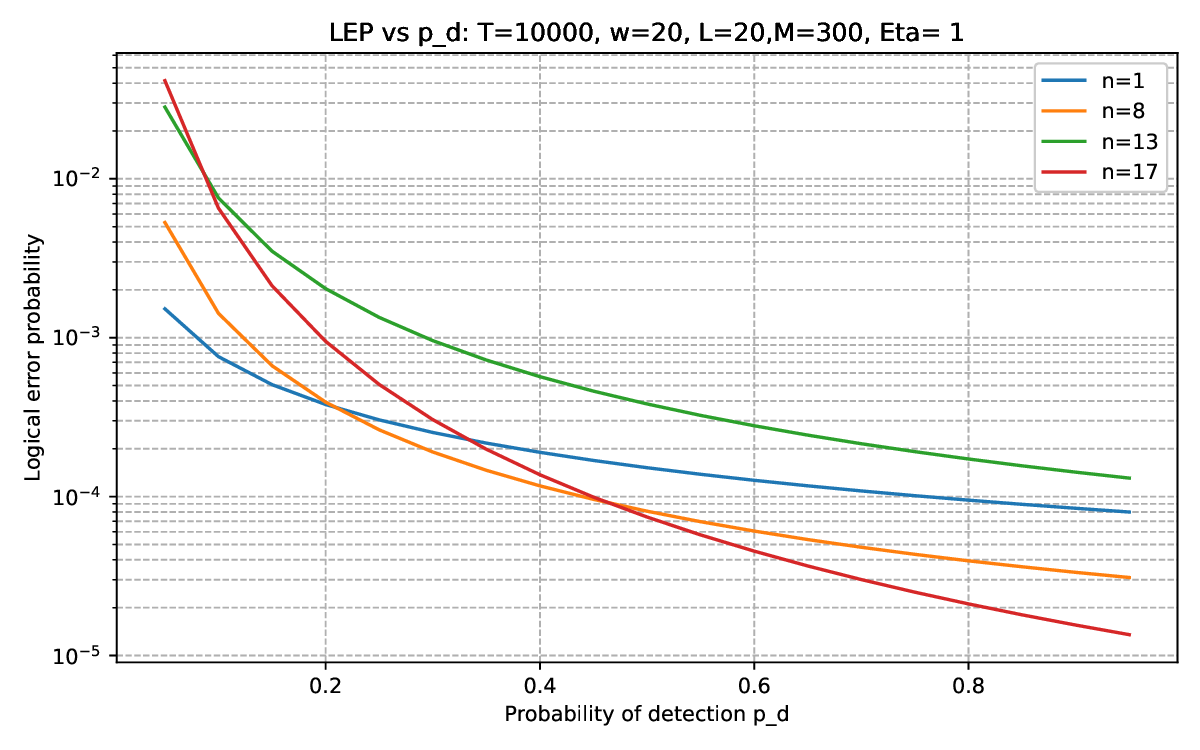}
\caption{Logical error probability $P_L$ as a function of detection probability $p_d$ for different puncturing tiers $n$ under symmetric errors.}
\label{fig:pl_vs_pd_symm}
\end{figure}

In addition to latency-constrained evaluations, we next examine how physical link quality affects logical reliability by studying the logical error probability $P_L$ as a function of the detection probability $p_d$, which enters the initial fidelity model in \eqref{eq:F0}. By evaluating $P_L$ as a function of $p_d$, we assess how different coding strategies perform under heterogeneous link conditions while maintaining the same underlying coding framework.

In Fig.~\ref{fig:pl_vs_pd_symm}, we show the logical error probability $P_L$ as a function of detection probability $p_d$ for symmetric errors under a fixed latency constraint $L=20$. At very low detection probabilities, corresponding to poor link quality or longer link distances, uncoded transmission achieves the lowest logical error probability, as coding provides limited error-correction benefit while increasing transmission error accumulation across multiple physical qubits. For moderate detection probabilities, approximately $0.2 \leq p_d \leq 0.4$, the $n=8$ punctured code achieves the lowest logical error probability, balancing error correction capability with reduced multi-qubit error accumulation relative to larger codes. Beyond approximately $p_d=0.5$, corresponding to improved link quality or shorter link distances, the $n=17$ code becomes favorable and achieves the lowest logical error probability. These decision regions show that different link conditions require different coding tiers, further motivating adaptive puncturing to accommodate heterogeneous network conditions.

\begin{figure}[t]
\centering    
\includegraphics[width=\linewidth]{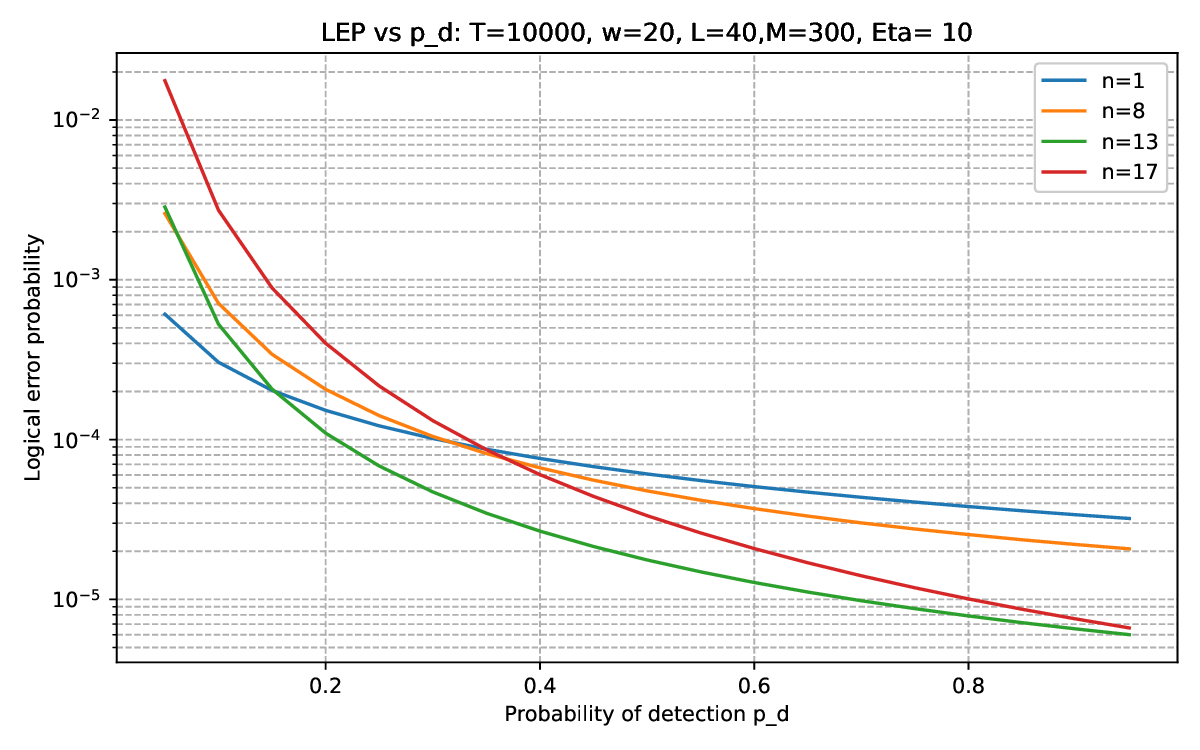}
\caption{Logical error probability $P_L$ as a function of detection probability $p_d$ for different puncturing tiers $n$ under asymmetric errors ($\eta=10$).}
\label{fig:pl_vs_pd_asymm}
\end{figure}

In Fig.~\ref{fig:pl_vs_pd_asymm}, we show that under asymmetric errors, the asymmetrically punctured $n=13$ code again plays a dominant role, achieving the lowest logical error probability over most of the detection probability range. While uncoded transmission performs best briefly at very low detection probabilities, the $n=13$ code becomes favorable once link quality improves, providing a significant reduction in logical error probability across most operating regimes. 

As in the latency-constrained results, these findings show that adaptive puncturing provides a flexible mechanism for selecting coding tiers under both symmetric and asymmetric error environments based on link conditions.

\section{Conclusions and Future Work}
In this work, we investigated the reliability of quantum teleportation under realistic network conditions where entanglement must be generated stochastically and degrades over time due to memory decoherence. Focusing on encoded teleportation, we developed a unified framework that captures the interaction between entanglement availability, decoherence, and coding decisions, enabling the evaluation of reliability in terms of logical error probability under latency constraints.

Our results demonstrate that encoded teleportation can provide substantial reliability gains over uncoded transmission under entanglement resource constraints. However, these gains depend critically on both the availability and fidelity of entangled pairs, as acquiring additional resources introduces delays that can reduce their effective quality. This interplay gives rise to non-uniform coding gains across operating regimes and highlights an interesting latency-reliability tradeoff. 

Through numerical investigations, we illustrated decision regions that determine  {optimized} effective code length under varying latency constraints {and link conditions}, and showed that code puncturing provides a practical mechanism for adapting the coding rate to resource availability without requiring changes to the underlying stabilizer structure. These findings underscore the importance of resource-aware adaptation in quantum networking, where both the quantity and quality of entanglement must be considered jointly in communication design.

Several directions for future work remain. First, while this work focuses on encoded teleportation, entanglement purification provides a complementary mechanism for improving resource quality; jointly optimizing purification and encoding strategies is a promising direction for further improving reliability. Second, extending the framework to dynamic settings where coding decisions are made online, based on the instantaneous availability of entangled pairs, would enable adaptive protocols that respond to time-varying resource conditions. Such extensions would further bridge the gap between theoretical models and practical quantum network implementations.

\bibliographystyle{IEEEtran}
\bibliography{bibliography}

\end{document}

%% file: Preamble.tex
\usepackage{cite}
\usepackage{amsmath,amssymb,amsfonts}
\usepackage{algorithmic}
\usepackage{graphicx, ifthen}
\usepackage{subfig}
\usepackage{tikz, pgfplots, pgfmath}
\usetikzlibrary{arrows.meta}
\usetikzlibrary{positioning, fit, calc, decorations.markings}
\usepackage{textcomp}
\usepackage{enumerate}
\usepackage{xcolor}
\usepackage{braket}
\usepackage{physics}
\usepackage{comment}
\usepackage{mathtools}
\usepackage{hyperref}
\usetikzlibrary{3d}
\usepackage{tikz-3dplot}
\usepackage{soul}
\usepackage{cancel}

\usepackage[stretch=15,shrink=20]{microtype}

\def\BibTeX{{\rm B\kern-.05em{\sc i\kern-.025em b}\kern-.08em
    T\kern-.1667em\lower.7ex\hbox{E}\kern-.125emX}}

\usepackage[utf8]{inputenc}
\usepackage[english]{babel}
\usepackage[T1]{fontenc}

\usepackage{lipsum}

\usepackage{makecell}
\setcellgapes{4pt}

\usepackage{amsthm}
\theoremstyle{plain}

\theoremstyle{definition}

\IEEEoverridecommandlockouts

%% file: Figures/figure_0.tex
\begin{figure*}[t]
\centering
\begin{tikzpicture}[
    x=1cm, y=1cm, >=latex, font=\sffamily,
    process/.style={draw=blue!70!black, fill=blue!5, thick, rounded corners=3pt, align=center, font=\small\sffamily, minimum height=0.8cm},
    usebox/.style={draw=orange!80!black, fill=orange!10, thick, rounded corners=3pt, align=center, font=\small\sffamily, minimum height=0.8cm},
    takeaway/.style={draw=teal!70!black, fill=teal!5, thick, rounded corners=4pt, align=center, font=\small\sffamily, text width=13.5cm, inner sep=10pt},
    branchtitle/.style={font=\large\sffamily\bfseries, text=black!80},
    note/.style={font=\footnotesize\sffamily, align=center, text=black!80},
    arrow/.style={->, >=latex, thick, draw=black!70},
    timeaxis/.style={->, >=latex, thick, draw=black!50},
    dimarrow/.style={<->, >=latex, thick, draw=black!60},
    waitline/.style={->, >=stealth, densely dashed, draw=red!70!black, thick},
    qubit/.style={circle, draw=blue!80!black, fill=cyan!20, thick, minimum size=0.45cm, inner sep=0pt}
]

\node[fill=white, font=\sffamily\scriptsize\bfseries, text=black!50] at (7.5, 9.3) {ENCODED TELEPORTATION};

\node[process, minimum width=3.5cm] (gen) at (2.5, 8.5) {Generate \& store\\EPR pairs};
\node[process, minimum width=2cm]   (enc) at (6.5, 8.5) {Encode};
\node[process, minimum width=2.5cm] (tel) at (9.5, 8.5) {Teleport};
\node[process, minimum width=2cm]   (dec) at (12.5, 8.5) {Decode};

\draw[arrow] (gen) -- (enc);
\draw[arrow] (enc) -- (tel);
\draw[arrow] (tel) -- (dec);

\draw[dimarrow] (0.75, 7.7) -- (4.25, 7.7) node[midway, below, note] {$t_{\mathrm{req}}$};
\draw[dimarrow] (5.5, 7.7) -- (7.5, 7.7) node[midway, below, note] {$t_{\mathrm{enc}}$};
\draw[dimarrow] (8.25, 7.7) -- (10.75, 7.7) node[midway, below, note] {$t_{\mathrm{tel}}$};
\draw[dimarrow] (11.5, 7.7) -- (13.5, 7.7) node[midway, below, note] {$t_{\mathrm{dec}}$};

\draw[timeaxis] (0, 6.9) -- (14.5, 6.9) node[right, font=\footnotesize\sffamily] {time};

\node[branchtitle] at (3.5, 6.4) {Case 1: Shorter Code ($[[n_1, k]]$)};

\draw[draw=black!60, fill=gray!10, thick, rounded corners=2pt] (0.5, 4.8) rectangle (3.7, 5.6);
\node[note, above] at (3.0, 5.6) {Quantum Memory (Storing $n_1$ entanglement resources)};

\node[process, minimum width=3.2cm] (gen1) at (2.1, 2.0) {Entanglement\\generation};
\node[usebox, minimum width=1.8cm]  (use1) at (5.5, 2.0) {Encode/ \\ Teleport/ \\Decode};
\draw[arrow] (gen1) -- (use1);

\foreach \i/\x in {1/0.9, 2/1.7, 3/2.5, 4/3.3} {
    \ifnum\i<4 \draw[gray!50, thick] (\x+0.4, 4.8) -- (\x+0.4, 5.6); \fi
    \node[qubit] (q1_\i) at (\x, 5.2) {};
    \draw[->, >=latex, thin, gray!80] (\x, 2.4) -- (\x, 4.8);
    
    \pgfmathsetmacro{\yline}{4.3 - \i*0.25}
    \draw[waitline] (\x, \yline) -- (4.6, \yline);
    \fill[red!70!black] (\x, \yline) circle (1.5pt); 
    \draw[thin, gray!50, dotted] (\x, 2.4) -- (\x, \yline);
}
\draw[thick, gray!70, dotted] (4.6, 2.4) -- (4.6, 4.4);
\node[font=\scriptsize\sffamily, text=red!80!black, left, align=right] at (4.5, 4.4) {decoherence due to wait times};

\draw[dimarrow] (0.5, 1.0) -- (3.7, 1.0) node[midway, below, note] {$t_{\mathrm{req}}(n_1)$};
\draw[timeaxis] (0, 0.4) -- (7, 0.4) node[right, font=\footnotesize\sffamily] {time};

\node[note, anchor=north] at (3.5, 0.2) {
    \begin{tabular}{l}
    \textbullet~Request completes sooner \\
    \textbullet~Less time in memory $\Rightarrow$ higher fidelity \\
    \textbullet~Lower coding gain, better entanglement
    \end{tabular}
};

\node[branchtitle] at (11.5, 6.4) {Case 2: Longer Code ($[[n_2, k]]$)};

\draw[draw=black!60, fill=gray!10, thick, rounded corners=2pt] (8.5, 4.8) rectangle (13.3, 5.6);
\node[note, above] at (10.9, 5.6) {Quantum Memory (Storing $n_2$ entanglement resources)};

\node[process, minimum width=4.8cm] (gen2) at (10.9, 2.0) {Entanglement\\generation};
\node[usebox, minimum width=1.8cm]  (use2) at (14.8, 2.0) {Encode/ \\Teleport /\\Decode};
\draw[arrow] (gen2) -- (use2);

\foreach \i/\x in {1/8.9, 2/9.7, 3/10.5, 4/11.3, 5/12.1, 6/12.9} {
    \ifnum\i<6 \draw[gray!50, thick] (\x+0.4, 4.8) -- (\x+0.4, 5.6); \fi
    \node[qubit] (q2_\i) at (\x, 5.2) {};
    \draw[->, >=latex, thin, gray!80] (\x, 2.4) -- (\x, 4.8);
    
    \pgfmathsetmacro{\yline}{4.5 - \i*0.22}
    \draw[waitline] (\x, \yline) -- (13.9, \yline);
    \fill[red!70!black] (\x, \yline) circle (1.5pt); 
    \draw[thin, gray!50, dotted] (\x, 2.4) -- (\x, \yline);
}
\draw[thick, gray!70, dotted] (13.9, 2.4) -- (13.9, 4.6);
\node[font=\scriptsize\sffamily, text=red!80!black, left, align=right] at (13.8, 4.4) {longer wait times};

\draw[dimarrow] (8.5, 1.0) -- (13.3, 1.0) node[midway, below, note] {$t_{\mathrm{req}}(n_2)$};
\draw[timeaxis] (8, 0.4) -- (16, 0.4) node[right, font=\footnotesize\sffamily] {time};

\node[note, anchor=north] at (11.5, 0.2) {
    \begin{tabular}{l}
    \textbullet~Request completes later \\
    \textbullet~More time in memory $\Rightarrow$ more decoherence \\
    \textbullet~Higher coding gain, lower entanglement
    \end{tabular}
};

\node[takeaway] at (8.0, -1.7) {
    \textbf{Implication:} $[[n_1,k]]$ is preferable in some regimes, while $[[n_2,k]]$ is preferable in others.\\[4pt]
    Support both using \textbf{punctured variants} of a common base code $[[n_{\mathrm{base}}, k]]$.
};

\end{tikzpicture}
\caption{Latency--reliability tradeoff in encoded quantum teleportation. The dominant latency arises from entanglement acquisition. With $n_1<n_2$, shorter effective code lengths require smaller entanglement packets, reducing acquisition burden and memory decoherence, but provide lower coding gain. Longer codes can provide stronger error protection, but require larger entanglement packets and may increase waiting times and storage-induced decoherence. Different latency--reliability regimes can therefore favor different code lengths, motivating adaptive puncturing from a common base code.}
\label{fig:latency_memory_tradeoff}
\end{figure*}

%% file: Figures/Figure_4.tex
\begin{figure}[t]
\centering
\makebox[\linewidth][c]{%
\begin{tikzpicture}[
    x=0.9cm, y=1cm, >=latex, font=\sffamily,
    curve1/.style={very thick, draw=blue!80!black},
    curve2/.style={very thick, dashed, draw=orange!80!black},
    curve3/.style={very thick, dotted, draw=green!60!black},
    boundary/.style={thick, densely dashed, draw=black!60},
    target/.style={thin, densely dotted, draw=black!60},
    lbl/.style={font=\small\sffamily, text=black!80},
    regionlbl/.style={font=\small\sffamily\bfseries, align=center}
]

\fill[blue!5]   (0,0) rectangle (2.2, 4.8);
\fill[orange!5] (2.2,0) rectangle (4.9, 4.8);
\fill[green!5]  (4.9,0) rectangle (8.2, 4.8);

\draw[->, thick, draw=black!80] (0,0) -- (8.6,0) node[right] {$L$};
\draw[->, thick, draw=black!80] (0,0) -- (0,5.1) node[above] {$P_L$};


\draw[curve1] plot [smooth, tension=0.6] coordinates { 
    (0.2, 3.6) 
    (2.2, 2.6) 
    (4.9, 2.4) 
    (8.2, 2.3) 
} node[right, text=blue!80!black] {$n_1$};

\draw[curve2] plot [smooth, tension=0.6] coordinates { 
    (0.2, 4.5) 
    (2.2, 2.6) 
    (3.55, 1.9) 
    (4.9, 1.5) 
    (8.2, 1.3) 
} node[right, text=orange!80!black] {$n_2$};

\draw[curve3] plot [smooth, tension=0.6] coordinates { 
    (0.2, 5.0) 
    (2.2, 3.8) 
    (4.9, 1.5) 
    (6.5, 0.9)
    (8.2, 0.6) 
} node[right, text=green!60!black] {$n_3$};

\draw[boundary] (2.2,0) -- (2.2,4.8);
\draw[boundary] (4.9,0) -- (4.9,4.8);

\draw[target] (0,3.2) -- (8.2,3.2);
\node[lbl, left] at (0,3.2) {$P_L^{(1)}$}; 

\draw[target] (0,1.9) -- (8.2,1.9);
\node[lbl, left] at (0,1.9) {$P_L^{(2)}$}; 

\draw[target] (0,0.9) -- (8.2,0.9);
\node[lbl, left] at (0,0.9) {$P_L^{(3)}$}; 

\node[regionlbl, text=blue!80!black]   at (1.1, 0.45) { $n^\star(L)=n_1$};
\node[regionlbl, text=orange!80!black] at (3.55, 0.45) { $n^\star(L)=n_2$};
\node[regionlbl, text=green!60!black]  at (6.5, 0.45) {$n^\star(L)=n_3$};

\node[lbl, below=3pt] at (2.2, 0) {$L_{12}$};
\node[lbl, below=3pt] at (4.9, 0) {$L_{23}$};

\end{tikzpicture}%
}
\caption{Decision regions induced by crossover behavior of the logical error probability $P_L$ under latency constraint $L$. Each curve corresponds to a code length $n$, and the preferred length $n^\star(L)$ minimizes $P_L$ at that latency. Larger $n$ requires a larger entanglement packet, changing the acquisition-time/fidelity tradeoff. The crossover points partition the latency axis into regions where different code lengths are preferred.}

\label{fig:decision_regions}
\end{figure}

%% file: Figures/age_Time.tex
\begin{tikzpicture}[
    x=1.55cm,
    y=1.12cm,
    font=\sffamily,
    >={Stealth[length=3mm]},
    line cap=round,
    line join=round
]

\path[use as bounding box] (-1.35,-0.9) rectangle (4.55,4.72);

\def\boxW{0.50}
\def\boxH{0.72}
\def\maxFill{0.90}

\definecolor{sucGreen}{RGB}{46,139,87}
\definecolor{failRed}{RGB}{205,92,92}
\definecolor{axisGray}{RGB}{70,70,70}
\definecolor{borderGray}{RGB}{120,120,120}
\definecolor{gridGray}{RGB}{215,215,215}

\tikzset{
mybox/.style={
draw=borderGray,
line width=0.8pt,
rounded corners=1pt
}
}

\newcommand{\fidelitybox}[4]{%
\begin{scope}[shift={(#1,#2)}]
\draw[mybox] (0,0) rectangle (\boxW,\boxH);

\fill[sucGreen,opacity=0.75]
(0.03,0.03) rectangle
({\boxW-0.03},
{0.03 + #4*\maxFill*(\boxH-0.06)});

\node[font=\tiny]
at ({0.5*\boxW},{0.58*\boxH})
{Age #3};

\end{scope}
}

\newcommand{\failurebox}[2]{%
\begin{scope}[shift={(#1,#2)}]

\draw[mybox,opacity=0] (0,0) rectangle (\boxW,\boxH);

\draw[failRed,line width=1.2pt,line cap=round]
(0.08,0.10) -- (\boxW-0.08,\boxH-0.10);

\draw[failRed,line width=1.2pt,line cap=round]
(0.08,\boxH-0.10) -- (\boxW-0.08,0.10);

\node[font=\tiny,text=failRed]
at ({0.5*\boxW},0.05)
{Failure};

\end{scope}
}

\foreach \x in {0.60,1.60,2.60,3.60}{
\draw[dashed,line width=0.65pt,gridGray]
(\x,-0.06)--(\x,4.62);
}

\foreach \y in {0.86,1.86,2.86,3.86}{
\draw[dashed,line width=0.65pt,gridGray]
(-0.82,\y)--(4.35,\y);
}

\draw[->,axisGray,line width=1.1pt]
(-0.82,-0.18)--(4.55,-0.18)
node[midway,below=6pt,font=\small\bfseries]
{Time step};

\draw[->,axisGray,line width=1.1pt]
(-0.82,-0.18)--(-0.82,4.75)
node[midway,left=5pt,rotate=90,font=\small\bfseries]
{Generation attempt};

\foreach \t in {0,1,2,3,4}{
\draw[axisGray,line width=0.8pt]
(\t+0.10,-0.12)--(\t+0.10,-0.22);

\node[below,font=\scriptsize]
at (\t+0.10,-0.22)
{$t=\t$};
}

\foreach \a/\y in {1/0.36,2/1.36,3/2.36,4/3.36,5/4.36}{
\draw[axisGray,line width=0.8pt]
(-0.76,\y)--(-0.88,\y);

\node[left,font=\scriptsize\itshape]
at (-1.10,\y)
{Attempt \a};
}

\fidelitybox{-0.15}{0}{0}{1.00}
\fidelitybox{0.85}{0}{1}{0.90}
\fidelitybox{1.85}{0}{2}{0.80}
\fidelitybox{2.85}{0}{3}{0.70}
\fidelitybox{3.85}{0}{4}{0.60}

\foreach \x in {0.85,1.85,2.85,3.85}{
\failurebox{\x}{1}
}

\fidelitybox{1.85}{2}{0}{1.00}
\fidelitybox{2.85}{2}{1}{0.90}
\fidelitybox{3.85}{2}{2}{0.80}

\foreach \x in {2.85,3.85}{
\failurebox{\x}{3}
}

\fidelitybox{3.85}{4}{0}{1.00}

\node[rotate=90,font=\tiny\bfseries,text=sucGreen!70!black]
at (-0.30,0.36)
{Fidelity};

\node[rotate=90,font=\tiny\bfseries,text=sucGreen!70!black]
at (1.70,2.36)
{Fidelity};

\node[rotate=90,font=\tiny\bfseries,text=sucGreen!70!black]
at (3.70,4.36)
{Fidelity};

\end{tikzpicture}

%% file: Figures/figure_5.tex
\begin{figure*}[t]
\centering
\begin{tikzpicture}[
    x=1cm, y=1cm, >=latex, font=\sffamily,
    data/.style={draw=black!70, fill=gray!5, thick, rounded corners=2pt, align=center, font=\footnotesize\sffamily, minimum height=1.0cm, inner sep=4pt},
    process/.style={draw=blue!80!black, fill=blue!5, thick, rounded corners=2pt, align=center, font=\footnotesize\sffamily, minimum height=1.0cm, inner sep=4pt},
    policy/.style={draw=teal!80!black, fill=teal!10, thick, rounded corners=2pt, align=center, font=\footnotesize\sffamily\bfseries, minimum height=1.0cm, inner sep=4pt},
    deploy/.style={draw=orange!80!black, fill=orange!10, thick, rounded corners=2pt, align=center, font=\footnotesize\sffamily, minimum height=1.0cm, inner sep=4pt},
    arrow/.style={->, thick, draw=black!70},
    groupbox/.style={draw=black!40, thick, dashed, rounded corners=4pt}
]


\fill[blue!2] (-0.2, 2.0) rectangle (15.5, 4.6);
\draw[groupbox, draw=blue!40] (-0.2, 2.0) rectangle (15.5, 4.6);
\node[anchor=north west, font=\small\sffamily\bfseries, text=blue!80!black] at (-0.2, 4.6) {Phase 1: Tradeoff Analysis \& Design};

\node[process, text width=2.8cm] (p1) at (1.8, 3.0) {Model Entanglement\\Generation vs.\\Latency $L$};

\node[process, text width=2.8cm] (p2) at (5.6, 3.0) {Analyze Stochastic\\Memory Decoherence\\\& Pauli Noise};

\node[process, text width=2.8cm] (p3) at (9.4, 3.0) {Evaluate $P_L$ vs. $L$\\ and $P_L$ vs. $p_d$\\ Tradeoffs across\\Puncturing Tiers $n$};

\node[policy, text width=2.4cm] (pol) at (13.5, 3.0) {Latency-Aware\\Decision Rule\\$n^\star(L, P_L)$};

\draw[arrow] (p1) -- (p2);
\draw[arrow] (p2) -- (p3);
\draw[arrow] (p3) -- (pol);


\fill[orange!2] (-0.2, -0.8) rectangle (15.5, 1.3);
\draw[groupbox, draw=orange!50] (-0.2, -0.8) rectangle (15.5, 1.3);
\node[anchor=north west, font=\small\sffamily\bfseries, text=orange!80!black] at (-0.2, 1.3) {Phase 2: Deployment Pipeline};

\node[data, text width=2.8cm] (req) at (1.8, 0.0) {Incoming Request:\\Latency $L_{\mathrm{req}}$ \&\\Target Rel. $P_L^{\mathrm{req}}$};

\node[deploy, text width=2.8cm] (enc) at (5.6, 0.0) {Determine $n^\star$,\\Puncture Base Code,\\\&\ Encode};

\node[deploy, text width=2.8cm] (tel) at (9.4, 0.0) {Teleport over\\Entangled Channel};

\node[deploy, text width=2.4cm] (dec) at (13.5, 0.0) {Decode \&\\Correct Errors};

\draw[arrow] (req) -- (enc);
\draw[arrow] (enc) -- (tel);
\draw[arrow] (tel) -- (dec);


\draw[->, >=latex, teal!80!black, dashed, very thick] 
    (pol.south) -- (13.5, 1.6) -- (5.6, 1.6) -- (enc.north);
    
\node[font=\footnotesize\sffamily, text=teal!80!black, fill=white, inner sep=2pt] 
    at (9.5, 1.6) {Inform Adaptive Code Selection};

\end{tikzpicture}
\caption{System framework for adaptive latency-constrained encoded teleportation. \textbf{Phase 1} represents the tradeoff analysis stage: we model entanglement generation under latency constraints, analyze the resulting stochastic memory decoherence, and evaluate the $P_L$ vs. $L$ and $P_L$ vs. $p_d$ tradeoffs to establish decision regions. \textbf{Phase 2} represents the physical deployment pipeline: incoming user requests specify constraints $(L_{\mathrm{req}}, P_L^{\mathrm{req}})$, which inform the code selection. The base code is punctured to the optimized length $n^\star$, {after which the data qubit is encoded,} followed by quantum teleportation, decoding, and error correction. In this work, we focus on phase 1 to motivate the need for adaptive coding strategies in the system. } 
\label{fig:system_model}
\end{figure*}